\documentclass[preprint,12pt]{elsarticle}

\usepackage{graphicx}
\usepackage{dcolumn}
\usepackage{bm}
\usepackage{amsfonts}
\usepackage{amsmath}
\usepackage{amssymb}
\usepackage{latexsym}
\usepackage{tikz}

\PassOptionsToPackage{normalem}{ulem}
\usepackage{ulem}
\providecolor{added}{rgb}{0,0,1}
\providecolor{deleted}{rgb}{1,0,0}

\journal{Applied Surface Science}

\begin{document}
\def\myfrac#1#2{\frac{\displaystyle #1}{\displaystyle #2}}

\begin{frontmatter}

\title{Wavelength Dependence of Picosecond Laser-Induced Periodic Surface Structures on Copper}

\author{Stella Maragkaki}
\ead{maragkaki@lat.rub.de}
\address{Chair of Applied Laser Technology, Ruhr-Universit\"at Bochum,
Universit\"atsstra\ss e~150, 44801 Bochum, Germany}

\author{Thibault J.-Y. Derrien}
\address{HiLASE Centre, Institute of Physics AS CR, Za Radnic\'i, 25241 Doln\'i B\u{r}e\u{z}any, Czech Republic}

\author{Yoann Levy}
\address{HiLASE Centre, Institute of Physics AS CR, Za Radnic\'i, 25241 Doln\'i B\u{r}e\u{z}any, Czech Republic}

\author{Nadezhda M. Bulgakova}
\address{HiLASE Centre, Institute of Physics AS CR, Za Radnic\'i, 25241 Doln\'i B\u{r}e\u{z}any, Czech Republic}
\address{S.S. Kutateladze Institute of Thermophysics SB RAS, 1 Lavrentyev ave., 630090 Novosibirsk, Russia}

\author{Andreas Ostendorf}
\address{Chair of Applied Laser Technology, Ruhr-Universit\"at Bochum,
Universit\"atsstra\ss e~150, 44801 Bochum, Germany}

\author{Evgeny L. Gurevich}
\ead{gurevich@lat.rub.de}
\address{Chair of Applied Laser Technology, Ruhr-Universit\"at Bochum,
Universit\"atsstra\ss e~150, 44801 Bochum, Germany}

\date{\today}

\begin{abstract}

The physical mechanisms of the laser-induced periodic surface structures (LIPSS) formation are studied in this paper for single-pulse irradiation regimes. The change in the LIPSS period with wavelength of incident laser radiation is investigated experimentally, using a picosecond laser system, which provides 7-ps pulses in near-IR, visible, and UV spectral ranges. The experimental results are compared with predictions made under the assumption that the surface-scattered waves are involved in the LIPSS formation. Considerable disagreement suggests that hydrodynamic mechanisms can be responsible for the observed pattern periodicity.
\end{abstract}


\end{frontmatter}


\section{Introduction}
Laser-induced periodic surface structures (LIPSS) or ripples were first reported by Birnbaum in 1965 \cite{Birnbaum} who attributed them to diffraction of the incident beam in the optical system. Nowadays two other theories are discussed in the context of the background mechanisms of the LIPSS formation (1) interference of the incident light with the surface scattered wave \cite{Sipe,Akhmanov} 
and (2) self-organization processes on the surface 
\cite{Costache,Varlamova,Reif,WhiteLight,Gurevich2016,Vilar}. 

In general, surface roughness allows for the coupling of the incident laser wave with surfaces that is described in the frames of the surface-scattered wave model \cite{Sipe}. For metallic/metalized surfaces, it is believed that surface scattered waves have the form of surface plasmon polaritons (SPP) \cite{Ursu} which turn to more localized plasmons with growing surface roughness \cite{Raether}. 
The influence of plasmonics is supported by the fact that, in many cases, the period of the observed LSFL (low-spatial frequency LIPSS) can be predicted in the frames of this model \cite{ThibaultJOptics}, and that the orientation
of the ripples follows the polarization of the incident light. 
Recent experiments on double-pulse generation of LIPSS
on silicon surfaces \cite{Thibault2}, 
as well as the LIPSS period reduction upon laser irradiation of silicon in water \cite{Thibault1}, are well explained in the frames of the plasmonic theory.

However, LIPSS generated by white light with a coherence length of $\sim$1.4\,$\mu$m cast some doubt on the purely interference hypothesis \cite{WhiteLight}\footnote{Calculation of the coherence length $\ell=\lambda_0^2/\Delta\lambda$ has been made based on the data of Fig. 3 in Ref. \cite{WhiteLight} with the central wavelength of the white light spectrum $\lambda_0\approx 650\,$nm and the line width $\Delta\lambda\approx 300\,$nm.}. Raman measurements reveal that the LIPSS formation takes place happens in the molten phase \cite{Reif}, which allows for hydrodynamic effects in the pattern formation. For multi-pulse irradiation regimes, it was concluded that the plasmonic stage, which governs the LIPSS orientation, does not necessary determine the periodicity of the final pattern due to contribution of the thermocapillary effects \cite{Colombier}. The hydrodynamic processes can be even more important for the LIPSS produced by single laser pulses at relatively high fluences assuming relatively deep melting and strong ablation of material \cite{Gedvilas,PRE}. Additionally, the hydrodynamic theory is supported by the topology of the pattern, which can be transient between the LIPSS and cell-like structures \cite{PRE}, that is typical for the hydrodynamic systems.

The accumulated evidences suggest that, in many situations, the plasmonic mechanism cannot completely explain the origin of LIPSS without assisting of other physical processes, especially for single laser pulses when the LIPSS emerge without progressive, pulse-by-pulse conversion of the surface roughness spectrum into periodic grating-like structure via gentle ablation \cite{Amoruso,VarlamovaAFM}. 
The main experimental argument supporting the plasmonic origin of LIPSS is
that the ratio between the period of the observed ripple pattern $\Lambda$ and the wavelength
of the incident light $\lambda$ fits the predictions of the plasmonic theory \cite{Bauerle}. However, to the best of our knowledge, the experimental evidences of this linear dependency are summarized for several wavelengths based on works of several independent groups, which generate the LIPSS patterns on different samples at different processing conditions such as laser fluence, repetition rate, focusing and pulse number.

In this paper, we report on systematic studies on the period of LIPSS generated by \textit{single} laser pulses of a picosecond laser system at three different wavelengths, of 1064, 532, and 355\,nm on the same copper sample. The experiments are described in Section 2. In Section 3, it is shown experimentally that, for the studied conditions, the observed LIPSS periodicity is considerably overestimated by the plasmonic theory. Possible mechanisms of this disagreement are discussed in Section 4.

\section{Experimental Setup}

\begin{figure}[h!]
 \centerline{\includegraphics[width=9cm]{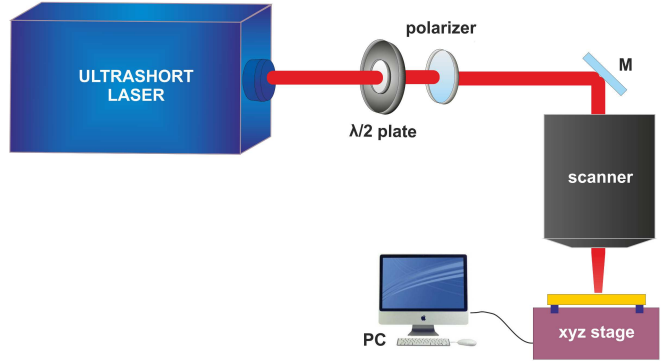}}
\caption{Experimental laser setup}
\label{setup}
\end{figure}

The LIPSS on the surface of a polished copper sample were generated at three different wavelengths with single pulses of a diode-pumped mode-locked picosecond laser (Lumera, Hyper Rapid 25, pulse duration $\tau_p\approx 7$\,ps, repetition rate set at 200\,kHz). The laser system was combined with a galvanometric scanner (SCANlab) equipped with an f-theta lens. The single pulse irradiation was achieved by high scanning speed of the galvoscanner controlled by a computer. The sample surface was positioned at the focal plane of the lens. The laser pulse energy was adjusted by $\lambda/2$-plate and a polarising beam splitter. 
The fundamental harmonics 1064\,nm, the frequency-doubled 532\,nm and the frequency-tripled 355\,nm wavelengths were used for the formation of periodic surface structures on Cu. All optical systems were identical, but consisted of elements designed for the corresponding spectral range. The one and the same scanner was used for the visible and the infrared light. A representative scheme of the laser setup used is shown in Fig.~\ref{setup}.

The structures were analyzed by means of optical and scanning electron microscopy. As one can see from the electron microscope images (Figs.~\ref{UV}-\ref{IR}), the LIPSS, which appear after single-pulse laser ablation are not ideally periodic and contain cell-like defects. These structures are less regular compared to multipulse-generated LIPSS. As a result, determination of the pattern period by Fourier transform of the whole image is ineffective and the LIPSS periods were determined through the pattern profile measurements. The number of peaks along the lines perpendicular to the ripples were measured in several defectless spots within the LIPSS-covered area for each laser wavelength. The peak fluence was calculated as $F= {2E_p}/{\pi\omega_0^2}$, where $\omega_0$ is the beam radius at the $1/e^2$ of the intensity and $E_p$ the laser pulse energy. We assume here that $M^2\approx 1$, i.e., that the pulses are nearly Gaussian.

\section{Experimental results}

\begin{figure}[h!]\centerline{\includegraphics[width=12cm]{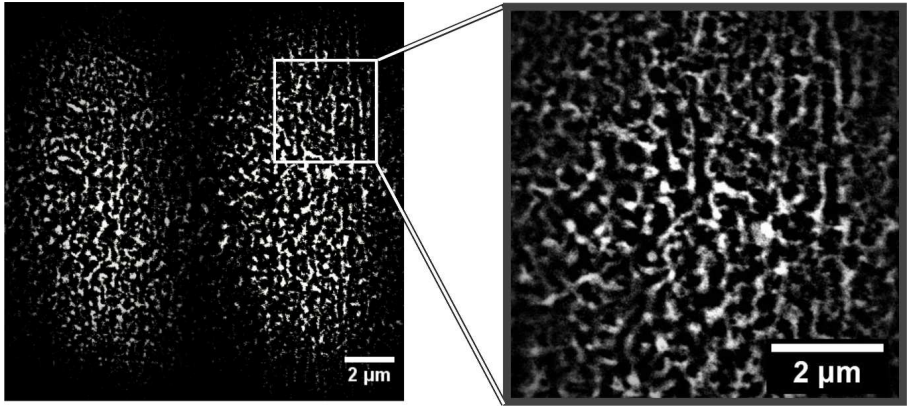}}
\caption{LIPSS on the copper surface exposed to single UV laser pulses; the wavelength $\lambda=355$\,nm, pulse duration $\tau_p\approx7$\,ps, peak fluence $F\approx 0.4\,\mathrm{J/cm^2}$. The average LIPSS period $\Lambda\approx 0.3\,\mathrm{\mu m}$. Two irradiation spots with centers separated by approximately 10\,$\mu$m are shown in the figure.}
\label{UV}
\end{figure}

\begin{figure}[h!]
\centerline{\includegraphics[width=12cm]{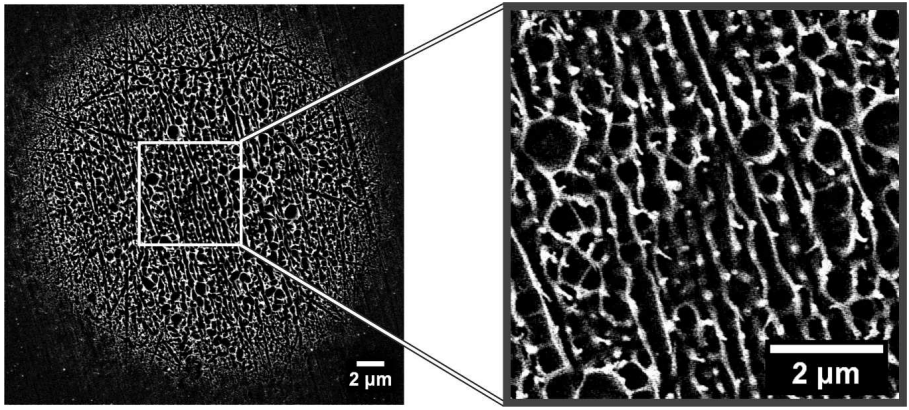}}
\caption{LIPSS on the copper surface exposed to single visible laser pulses; $\lambda= 532$\,nm, $\tau_p\approx7$\,ps, $F\approx 3.5\,\mathrm{J/cm^2}$. The average LIPSS period $\Lambda\approx 0.36\,\mathrm{\mu m}$.}
\label{greenps}
\end{figure}

\begin{figure}[h!]
\centerline{\includegraphics[width=12cm]{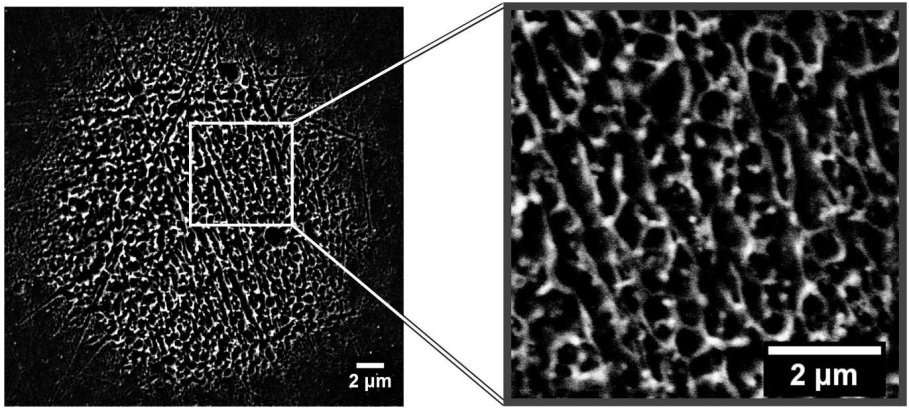}}
\caption{LIPSS formed on copper exposed to single IR laser pulses; $\lambda= 1064$\,nm, $\tau_p\approx7$\,ps, $F\approx 5.5\,\mathrm{J/cm^2}$. The average LIPSS period $\Lambda\approx 0.58\,\mathrm{\mu m}$.}
\label{IR}
\end{figure}

In this work, we investigate the wavelength dependence of the periodicity of LIPSS produced by single picosecond laser pulses, making an attempt to correlate it with the plasmonic theory and gain a better insight into the underlying physical mechanisms. The plasmonic concept of LIPSS formation was never systematically tested before for the same metal sample irradiated under the same conditions by laser pulses of different duration. Figures \ref{UV}--\ref{IR} show the electron microscope images of LIPSS produced by single picosecond laser pulses at different wavelengths. In each Figure, laser fluence corresponds approximately to the middle of the range, in which the LIPSS can be observed at particular wavelengths.

Table~\ref{summary} summarizes the obtained theoretical and experimental data on the LIPSS periods. Considerable disagreement of the plasmonic theory with experiment, which increases with laser wavelength and cannot be explained by both the statistical and the systematic error, calls for revision of the LIPSS formation concept. 
\begin{table}
\begin{tabular}{c|c c|c}
\hline \hline
$\lambda$ (nm) & $\Lambda_t^{(1)}$ (nm) & $\Lambda_t^{(2)}$ (nm) &$\Lambda_{exp}\pm\sigma$ (nm) \\
\hline
1064 & 1049 & 1037 & {580} $\pm$ {70} \\
532 & 509 & 508 & {360} $\pm$ {20} \\
355 & 355 & 330 & {300} $\pm$ {40} \\
\hline
\hline
\end{tabular}

\caption{Summary of the experimental and theoretical results. $\lambda$ is the incident light wavelength. $\Lambda_t^{(1)}$ is the LIPSS period calculated using the simple SPP model (solid line in Fig.~\ref{Lcalc} ). $\Lambda_t^{(2)}$ is the LIPSS period calculated with accounting for the surface roughness (squares in Fig.~\ref{Lcalc}). $\Lambda_{exp}$ is the experimentally measured LIPSS period. $\sigma$ is the measurement uncertainty.} 
\label{summary}
\end{table}

\section{Discussion}
\begin{figure}[h!]
 \centerline{
 \includegraphics[width=10cm]{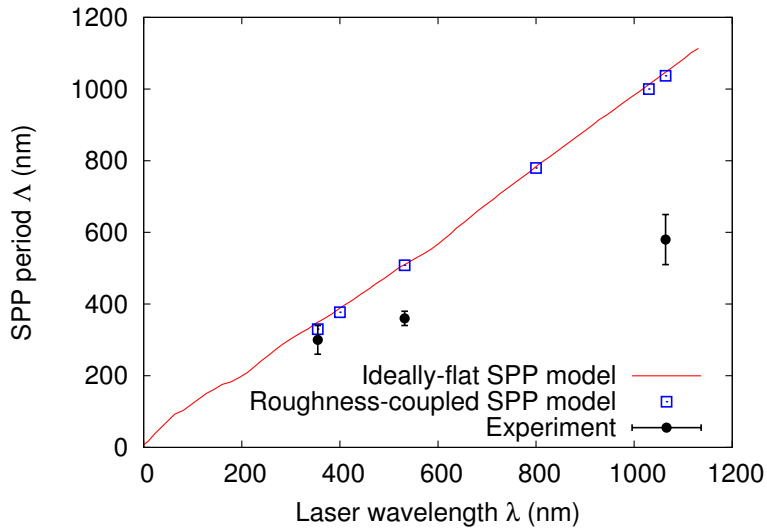}}
\caption{LIPSS period for copper as a function of laser wavelength. Two models are compared. Red solid line was calculated by Eq.~\eqref{Lambda} using the data on optical properties from~\cite{Palik1985}. Empty squares were calculated accounting for the surface roughness as summarized in~\cite{Bonse2005}. Experimentally measured LIPSS periods $\Lambda$ are shown by dots with error bars.
}
\label{Lcalc}
\end{figure}

In the frames of the plasmonic (or interference) model, the spatial period of the LIPSS on metallic surfaces $\Lambda$ depends on the wavelength of the incident light $\lambda$ as \cite{Bauerle}
\begin{equation}
\Lambda = \frac{\lambda}{\mathcal{R}\!e\left(\sqrt{\frac{\varepsilon}{1+\varepsilon}}\right)}
\label{Lambda} 
\end{equation}
where the $\varepsilon$ is the wavelength-dependent dielectric constant of metal. The angle of the laser beam incidence is equal to zero for all experiments reported here. 
The LIPSS period for copper calculated according to Eq.~\eqref{Lambda} with the data on optical properties from \cite{Palik1985} is shown in Fig.~\ref{Lcalc} by solid line.

The simple SPP model, expressed by Eq.~\eqref{Lambda}, assumes an ideally-flat sample surface. Interestingly, the calculations based on a more advanced model (so-called Sipe model \cite{Sipe}), which introduces the shape and filling factors to describe randomly rough surfaces, $s$ and $f$ respectively, coincide with predictions of the simple SPP theory, Eq. (1), with $s=0.4$ and $f=0.1$ (as given in Ref. \cite{Bonse}). The LIPSS period predicted by this model is given in Fig.~\ref{Lcalc} by empty squares. The experimental data shown by dots with error bars demonstrate that, for the conditions of LIPSS formation studied here, both the simple and sophisticated models can predict only the overall tendency of LIPSS period evolution with laser wavelength but cannot describe the period values. For stainless steel surfaces, Gedvilas et al. \cite{Gedvilas} have also reported periodicities of LIPSS created by single picosecond pulses, which are considerably smaller than predicted in the frames of the plasmonic theory. Note that Equation~\eqref{Lambda} yields respectively $\sim$350\,nm and $\sim$1050\,nm at 355 and 1064\,nm wavelengths for optical properties of iron \cite{Palik1985}. 

Two possible interpretations of the disagreement are possible: 

(1) During picosecond laser pulses, the optical constants of metals can considerably change due to swift heating of the conduction electrons followed by heat exchange with the lattice and lattice melting. Although at picosecond irradiation regimes of moderate fluences the non-equilibrium between the electron and lattice subsystems is not so strong as in the case of femtosecond laser pulses (see, e.g., \cite{kirkwood_experimental_2009_PRB}), the simulations for the present experimental conditions based on the two-temperature model (TTM) \cite{Yoann2016} reveal extremely high heating of the lattice already by  the middle of the laser pulse. Thus, at $\lambda$ = 355\,nm, the laser pulse with peak fluence of 0.4\,J/cm$^2$  (Fig.~\ref{UV}) corresponding to the fluence in the center of irradiation spot yields electron and lattice temperatures in the excess of 5000\,K. Under the irradiation conditions of Figs.~\ref{greenps} and~\ref{IR}, modelling gives lattice heating up to 30 000\,K. Although simulations of such regimes, when matter experiences heating to extreme states and can behave abnormally \cite{Cho}, calls for a more sophisticated modeling, the TTM simulations unambiguously show the tendency for enormous heating of matter. 

Such a level of heating assumes material melting with possible superheating of the solid phase \cite{Zhigilei}, initiation of lattice disintegration already during first picoseconds \cite{Zhigilei,Semerok,Povarnitsyn}, formation of stress and unloading waves in the surface layer of the sample \cite{Zhigilei,Povarnitsyn}. All these processes should lead to the transient changes of the dielectric function, so that the room-temperature optical parameters \cite{Palik1985} are not applicable anymore for estimating of the LIPSS period. However, the evolution of optical properties of metallic surfaces dynamically evolving at picosecond timescales is poorly known while the concept of plasmons on such surfaces requires reconsideration. We keep this explanation as possible, but the detailed analysis of this possibility would require development of a new theory of the material optical response upon interaction between ultrashort laser pulses and metals that is beyond the scope of this paper.

\begin{figure}
\begin{tikzpicture}
\draw[thick,->](0,0)--(0,3)node[right]{$T_e$};
\draw[thick,->](0,0)--(3,0)node[above]{$x$};
 \draw[thick,color=blue,domain=0:9,smooth]plot (\x/3,{0.3*sin(80*\x)+2});
   \draw[thick,->](4,0)--(4,3)node[right]{$T_l$};
   \draw[thick,->](4,0)--(7,0)node[above]{$x$};
   \draw[thick,color=red,domain=0:9,smooth]plot (\x/3+4,{1.3*sin(80*\x)+1.5});
   \node at (9.25,2.1){\includegraphics[width=2.5cm]{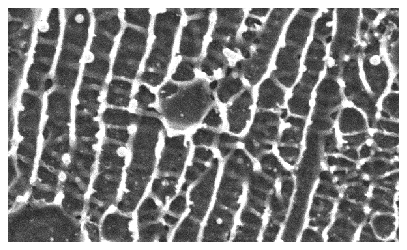}};
   \node at (9.25,0.7){Hydrodynamic};
   \node at (9.25,0.1){Instabilities};
   \draw[line width=5pt,green,->](1,3.2)..controls(1,4)..(2,4);
   \node at (3,4){\includegraphics[width=2cm]{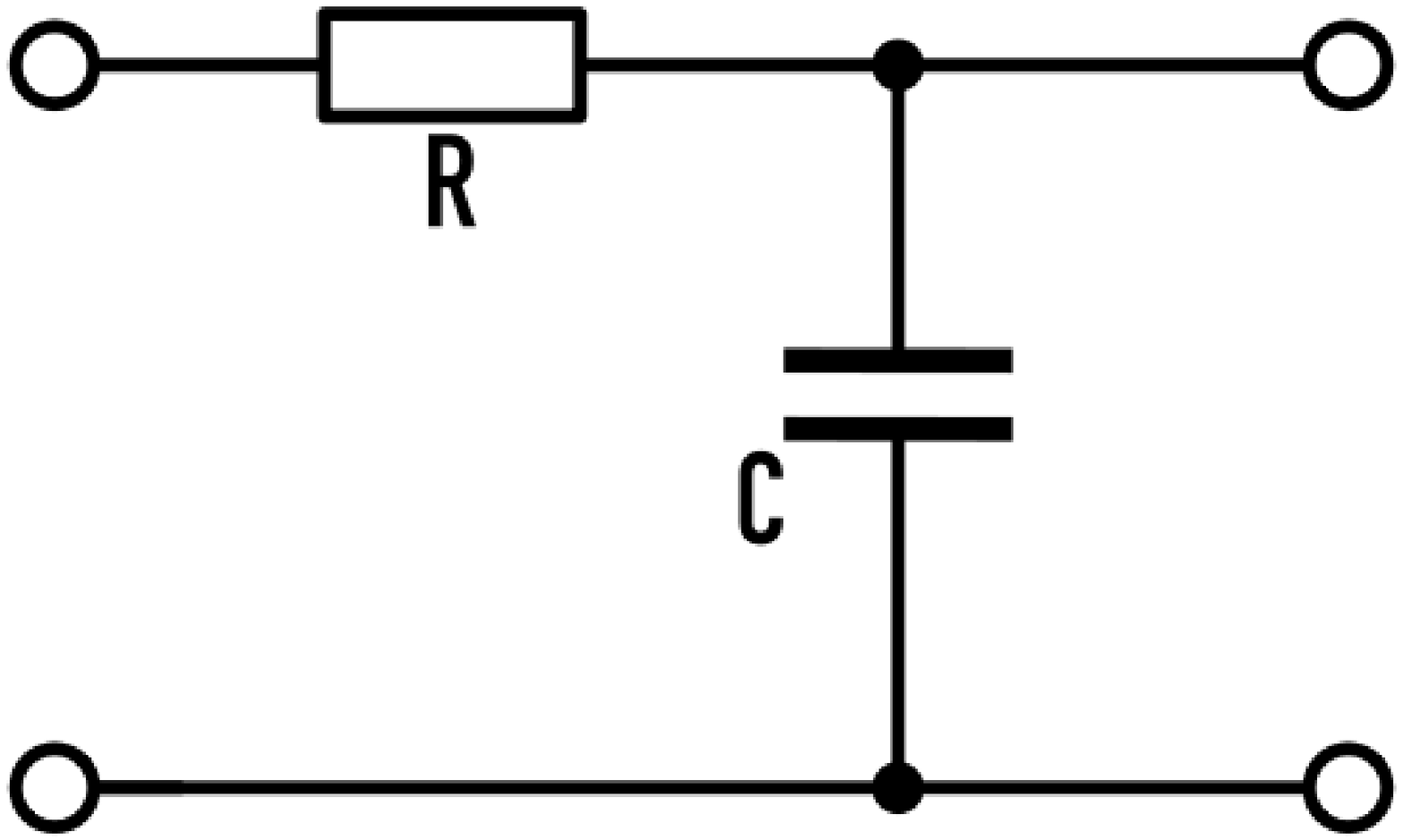}};
   \draw[line width=5pt,green,->](4,4)..controls(5,4)..(5,3.2);
   \draw[line width=5pt,green,->](6,3.2)..controls(6,4)..(7,4);
   \node at (8,4){\includegraphics[width=2cm]{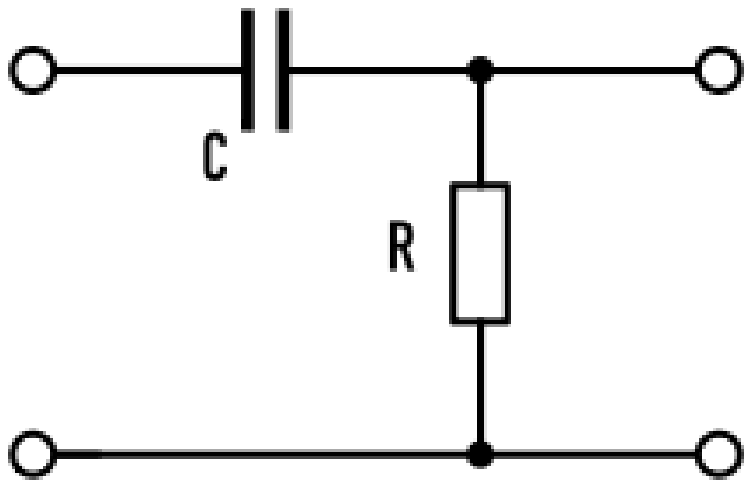}};
   \draw[line width=5pt,green,->](9,4)..controls(10,4)..(10,3.2);
\end{tikzpicture}
\caption{Schematic representation of the three-step model. LIPSS are generated as a result of three consecutive steps: electron temperature modulation $\rightarrow$ lattice temperature modulation $\rightarrow$ surface pattern formation. Efficiency of each step depends on the wavelength of the preceding modulation.}\label{3steps}
\end{figure}

(2) The second possible explanation, on which we concentrate in this paper, is based on the the three-step model of the LIPSS formation:
\begin{enumerate}
\item Periodically modulated electron temperature can appear due to, e.g., interference of the incident light and the surface scattered waves or periodic modulation of the incident light which in its turn causes a modulated reflection of the sample surface. 
\item The amplitude modulation in the electron temperature evolves into a high-amplitude modulation in the lattice temperature upon electron-lattice thermalization. This process was studied in detail in \cite{Yoann2016,PRB_instability}. 
\item Hydrodynamic processes in the molten surface layer relocate material along the surface and form the LIPSS pattern. Possible hydrodynamic instability mechanisms are discussed in \cite{Varlamova,Reif,Gurevich2016,AcuosticWave2}
\end{enumerate}
Schematically this model is represented in Fig.~\ref{3steps}.

We stress that the spectrum of the initial modulation of the electron temperature $T_e$ is rather broad, because the spectrum of the surface roughness is not monochromatic. Moreover, the incident beam intensity is non-uniformly distributed over the irradiation spot on the sample surface (e.g., due to the Gaussian beam profile) that can also lead to broadening of the spectrum of the $T_e$ modulation. The efficiency of each aforementioned step is higher or lower for different spatial periods of the instabilities, thus acting as high-pass or low-pass filters. In the frames of this 3-step model, according to our experimental observations, the complete process of the LIPSS formation should be a high-pass filter, because it works better for high spacial frequencies of the initial temperature modulation (in the UV and visible ranges), whereas for low periods (in the IR range) the observed LIPSS period is considerably smaller than predicted by the plasmonic theory. 

It has been demonstrated \cite{PRB_instability}, that the transition between the steps 1 and 2 acts as a low-pass filter, because the growth rate of the lattice temperature modulation with larger spatial periods is higher, than that of the high-frequency temperature modulations. Consequently, in order to explain the experimental observations, the growth rate of the hydrodynamic instability must decrease with the period, i.e., the instability should work as a high-pass filter (or a band-pass filter with the central frequency in the short-wave range). 
Our estimating TTM simulations show that, for the three cases considered, the ablation depths are in the range of 100--300\,nm leaving behind a molten metal layer app. 50\,nm thick.
Material relocation within this layer induced by hydrodynamic instability, in particular at post-irradiation timescales, should be dependent on many factors, including molten layer depth, melt viscosity and surface tension. These factors can provide the reasons for the deviation of the LIPSS period from the initial interference pattern imprinted on the surface by the interference of incident and scattered electromagnetic waves.

\section{Conclusions}

In this paper, it is shown that the periodicity of the LIPSS patterns on metal surfaces at single-pulse irradiation regimes can be explained by the mutual action of the two known mechanisms, laser light interference with the scattered electromagnetic wave, which is responsible for the LIPSS direction, and consequent relocation of molten material facilitated by the hydrodynamic instability of melt, which finally affects the pattern periodicity. Additional limitations on possible hydrodynamic instabilities, which participate in the melt relocation on the irradiated surface are discussed. The instability should act as a high-pass filter, i.e., the growth rate should increase at shorter periods of the initial surface temperature modulations. As an example, the growth rate of the ablative Rayleigh-Taylor instability is proportional to $\Lambda^{-1/2}$ and, hence, satisfies this condition. The temperature-modulated ablation efficiency \cite{Gurevich2016} can also be considered as a possible mechanism of the LIPSS formation. For a given amplitude of the lattice temperature modulation on the surface, the difference between the driving force of this instability and the viscous friction is larger for smaller periods. Due to this, the final LIPSS period is to be shifted to shorter wavelengths and this mechanism can be considered as a high-pass filter.

\section{Acknowledgment}

TJYD, YL and NMB acknowledge the support from the European Regional Development Fund and the state budget of the
Czech Republic (project BIATRI: CZ.02.1.01/0.0/0.0/15$\_$003/0000445) and from the Ministry of Education, Youth and Sports of the Czech Republic (Programs NPU I project no. LO1602, and Large Research Infrastructure project no. LM2015086).  TJYD also acknowledges funding from the European Commission for the Marie Sklodowska-Curie Individual Fellowship (project 657424).
\bibliographystyle{elsarticle-num}

\end{document}